%% file: Claude.tex
\documentclass[a4paper,twoside,11pt,openright,titlepage]{mybook}  
\usepackage[latin1]{inputenc}
\usepackage[T1]{fontenc}
\usepackage{float}
\usepackage[fleqn,reqno]{amsmath}
\usepackage{amsfonts,amssymb,amsxtra}
\usepackage{ae}
\usepackage{lscape}
\usepackage{url}
\usepackage{multirow}
\usepackage[usenames]{color}

\usepackage{natbib}
\usepackage[german,english]{babel}
\usepackage{scalefnt}

\usepackage{epic,eepic}

\usepackage{vmargin}
\setpapersize[portrait]{A4}
\normalsize
\setmarginsrb{2.4cm}{2.0cm}{2.4cm}{3.0cm}%
             {1\baselineskip}{1.5\baselineskip}{1\baselineskip}{2\baselineskip}

\newcommand{\clearemptydoublepage}{\newpage{\pagestyle{empty}\cleardoublepage}}

\newcommand{\bq}{\begin{eqnarray}}
\newcommand{\eq}{\end{eqnarray}}                                   

\usepackage{fancyhdr}
\renewcommand{\chaptermark}[1]{\markboth{\thechapter. #1}{}}

\usepackage{epsfig}
\usepackage{floatflt}     
\usepackage{subfigure}    

\usepackage[bf]{caption}  
\usepackage[bookmarks=true]{hyperref}
\begin{document}
\selectlanguage{english}
\frontmatter
\pagenumbering{roman}
\pagestyle{empty}

\clearemptydoublepage

\include{title}

\clearemptydoublepage

\lhead[\fancyplain{}{\sl \thepage}]%
 {\fancyplain{}{\sl Contents}}
\rhead[\fancyplain{}{\sl Contents}]%
 {\fancyplain{}{\sl \thepage}}

\tableofcontents
\addcontentsline{toc}{chapter}{Contents}
\clearemptydoublepage

\mainmatter
\pagenumbering{arabic}
\pagestyle{fancyplain}
\cfoot{}

\lhead[\fancyplain{}{\sl \thepage}]%
 {\fancyplain{}{\sl \rightmark}}
\rhead[\fancyplain{}{\sl \leftmark}]%
 {\fancyplain{}{\sl \thepage}}


\include{introduction}


\include{usingclaude}


\include{siteinformation}


\include{images}


\include{acquisition}


\include{flatwizard}


\include{scheduler}


\include{options}


\pagenumbering{Roman}

\lhead[\fancyplain{}{\sl \thepage}]%
 {\fancyplain{}{\sl Bibliography}}
\rhead[\fancyplain{}{\sl Bibliography}]%
 {\fancyplain{}{\sl \thepage}}
\addcontentsline{toc}{chapter}{Bibliography}
\bibliographystyle{doktor_natb_bab_amp}

\parindent 0.0em

\bibliography{dissbib}
\clearemptydoublepage


\lhead[\fancyplain{}{\sl \thepage}]%
 {\fancyplain{}{\sl Appendix}}
\rhead[\fancyplain{}{\sl Appendix}]%
 {\fancyplain{}{\sl \thepage}}
\include{appendix}
\clearemptydoublepage


\end{document}

%% file: title.tex
\begin{flushleft}

{\scshape \Large Tim-Oliver Husser} \\
{\small Institut f\"ur Astrophysik G\"ottingen\\
husser@astro.physik.uni-goettingen.de}

\vspace*{1cm}
{\scshape {\Huge Claude \\[6pt]}
  {\Large An Automation Tool for the Monet Telescopes}
} 

\vspace*{1.5cm}
{\scshape \Large Manual}\\
(\today)

\begin{figure}[!hb]
  \centering
  \vspace*{1cm}
  \includegraphics[width=\textwidth]{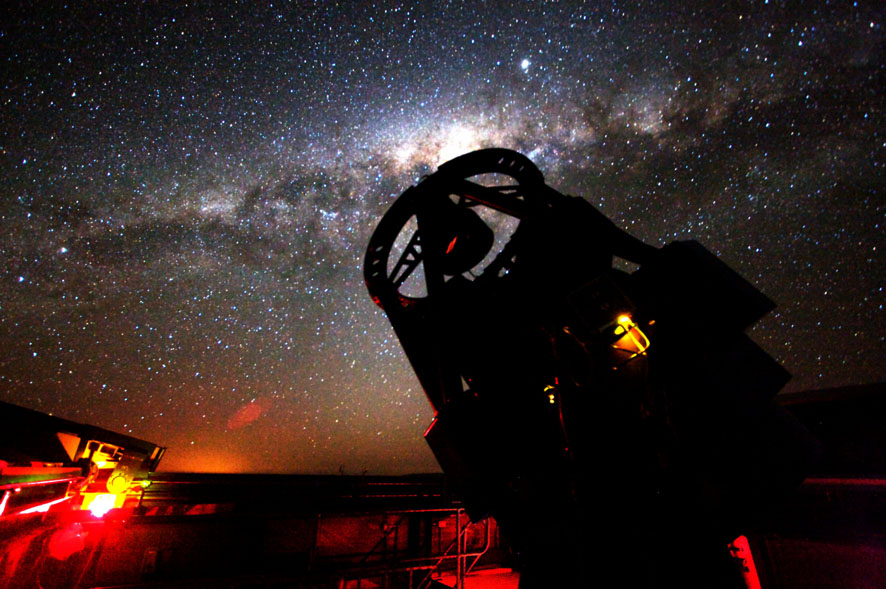}
\end{figure}

\end{flushleft}

\newpage
\color{white}platzhalter\color{black}
\vspace{18cm}
\begin{tabbing}
  ~~~~~~~~~~~~~~~~~~~~~~~~~~~~~~~~ \hspace{4.5cm} \= \\
  Image: Stephen Potter, SAAO, 2008\\ 
\end{tabbing}

%% file: introduction.tex
\chapter{Introduction}

\section{Monet}
MONET (MOnitoring NEtwork of Telescopes) is a telescope project of the 
Georg-August-Universit\"at G\"ottingen, the University of Texas at Austin and the South
African Astronomical Observatory (SAAO). Two identical telescopes have been placed
at the McDonald Observatory (2005) in Texas, and at the SAAO site (2008) in Sutherland, South
Africa.

The telescopes have been funded by the 
\emph{Alfried Krupp von Bohlen und Halbach-Stiftung}\footnote{\url{http://www.krupp-stiftung.de/}} and
built by Halfmann Teleskoptechnik\footnote{\url{http://www.halfmann-teleskoptechnik.com/}}.

While 50 per cent of the observation time is used for science observations by the three partners,
the rest of the time is dedicated to be used by schools for the education of their students. This is
made possible by a piece of software that allows the observer to fully control the telescope over
the internet via a simple web browser. Due to the time shift between the telescope and potential 
observers, e.g. students in Germany can use the telescope in Texas during normal school hours, while
it is still night there.

Furthermore the locations of the two telescopes allow us to cover the whole sky, i.e. every bright enough
object in the universe is observable by the \emph{Monet} telescopes. Also it is possible to start the
observation of objects near the cellestial equator in South Africa until sunrise and then continue using the
telescope in Texas.

The \emph{Monet} telescopes were designed to work in a robotic mode, i.e. without any observer interaction.
A list objects to be observed would then be sent to the telescope and everything would be done
automatically. Unfortunately that mode is not available yet.

\section{The Monet portal \& Claude}
The \emph{Monet portal} is a web page that allows logged in users to control the \emph{Monet} telescopes.
The telescope can be moved to any coordinates accessible from the telescope's position and pictures can be
taken, both using a simple user interface.

While this is completely adequate for simple observations and helps getting students at schools
working with the telescopes, especially professional astronomers are missing some more
advanced functionality. I.e. focussing the telscope is a task that can be automated quite easily, but
with using only the \emph{Monet portal} it means subsequently changing the focus value and taking a
picture. And in the end the observer has to fit the data manually to get the best focus. \emph{Claude}
has been designed to help with this and in general automate standard procedures as good as possible.

\emph{Claude} is not a replacement but more like an extension of the \emph{portal}. Basically it has a web
browser built in, which does nothing else than displaying the \emph{Monet portal}. But due to this,
\emph{Claude} is able to use the \emph{portal} as a normal user would, i.e. fill in forms and click
on buttons. This allows for some quite advanced automating of the \emph{Monet} telescopes.

Furthermore \emph{Claude} tries to cover everything an observer needs to know about the telescopes, i.e.
it shows a list of recent images (and offers to download them), shows weather informations and automatically
fills in a night log for the observer.

At the moment \emph{Claude} is only used internally at the \emph{Institut f\"ur Astrophysik G\"ottingen}, but
at some point we are planning to make it public, so that it can be used by every \emph{Monet} observer,
especially schools.

%% file: usingclaude.tex
\chapter{Using Claude}

\section{First start}
When \emph{Claude} is started for the first time, the options dialog (see chapter \ref{chap.options}) is shown.
The only option that really needs to be set now is the \textbf{Claude dir}. \emph{Claude} will not start
without this being set properly.

\section{Wrapping the Monet portal}
\label{sec.WrappingPortal}
\begin{figure}
  \centering
    \includegraphics[width=\textwidth]{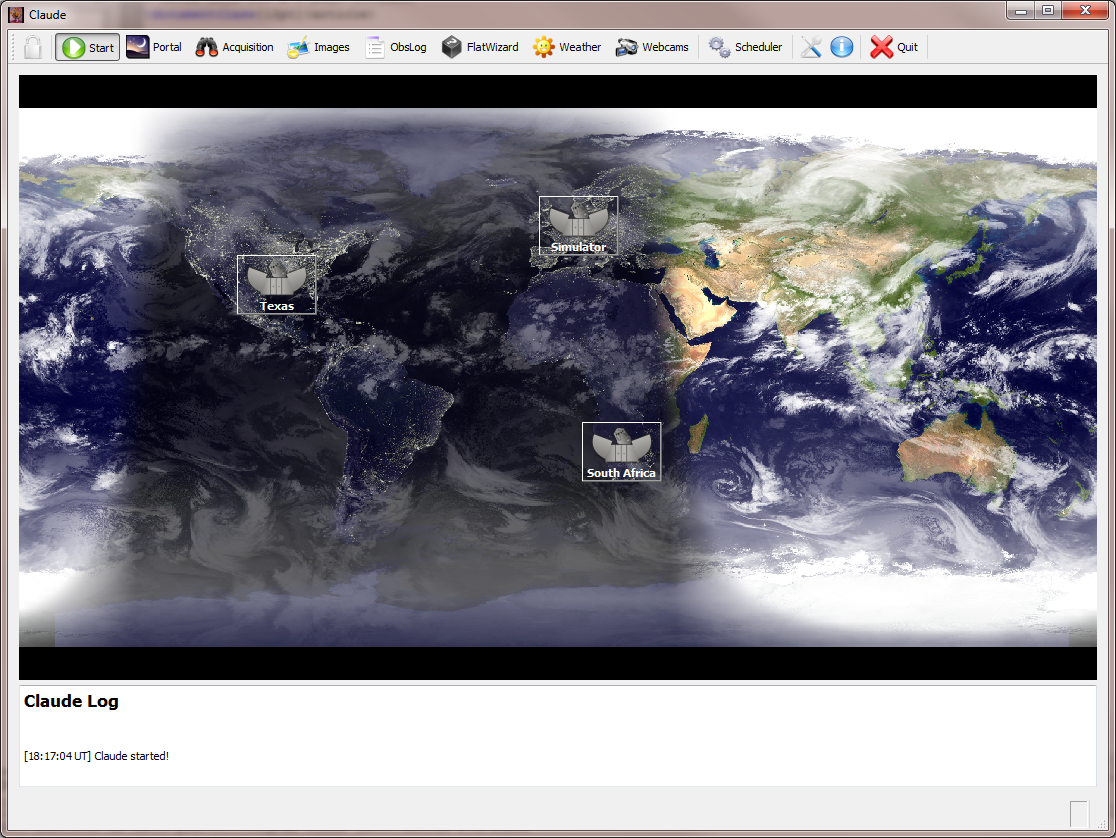}
  \caption{The start screen of Claude.}
  \label{fig.StartScreen}
\end{figure}
After starting Claude the first thing one sees is a world map (see fig. \ref{fig.StartScreen}) 
with three buttons on it for \emph{Monet/North} in Texas, \emph{Monet/South} in South Africa and 
the \emph{Monet Simulator}. The bright and dark areas on the map depict the current day/night 
distribution on earth, so you can always check which of the telescopes should see some stars above. 
The earth at night is composed of a \emph{NASA} image\footnote{\url{http://visibleearth.nasa.gov/view_rec.php?id=11793}},
while the daylight image is provided by Tor {\O}ra\footnote{\url{http://www.oera.net/How2/TextureMaps.htm}}.

The world map also presents the actual cloud coverage currently measured on Earth and is updated 
every thirty minutes. The current cloud map is taken from the \emph{Xplanet project}
\footnote{\url{http://xplanet.sourceforge.net/}}.

\begin{figure}
  \centering
  \subfigure[Start page of Monet/North portal.]{\includegraphics[width=0.48\textwidth]{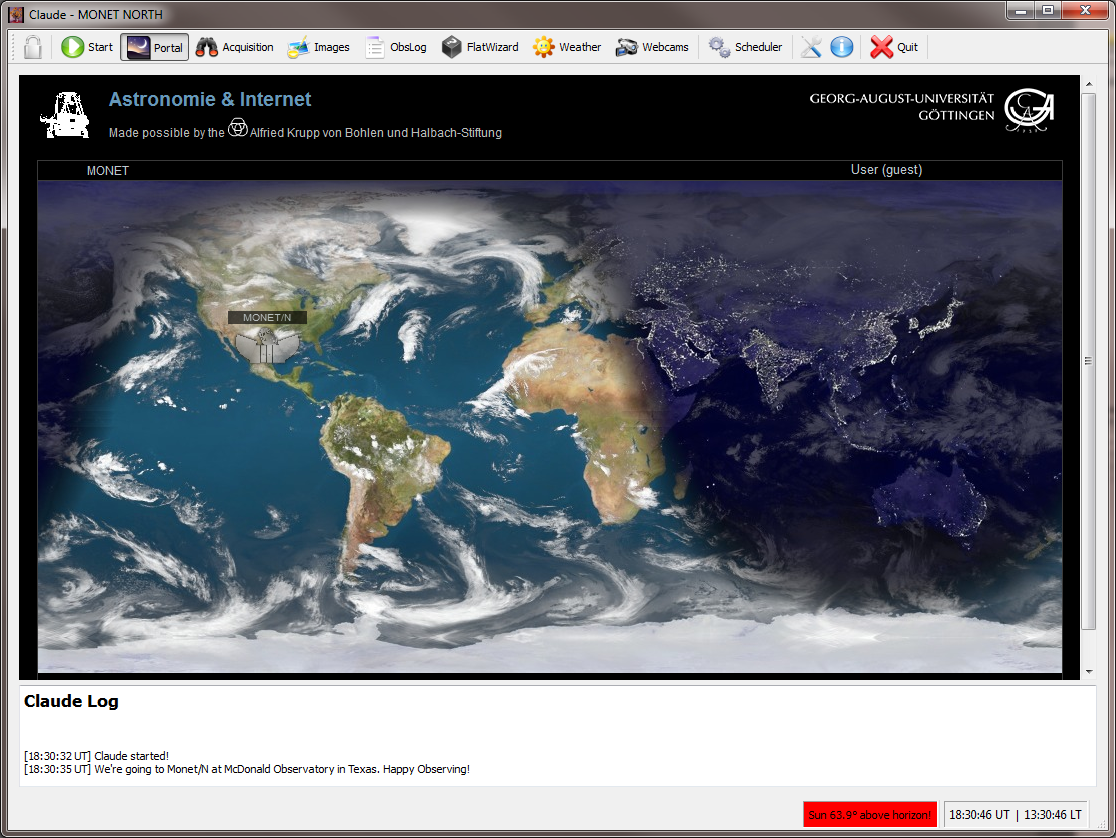}}           
  \subfigure[Main page of Monet/Sim portal.]{\includegraphics[width=0.48\textwidth]{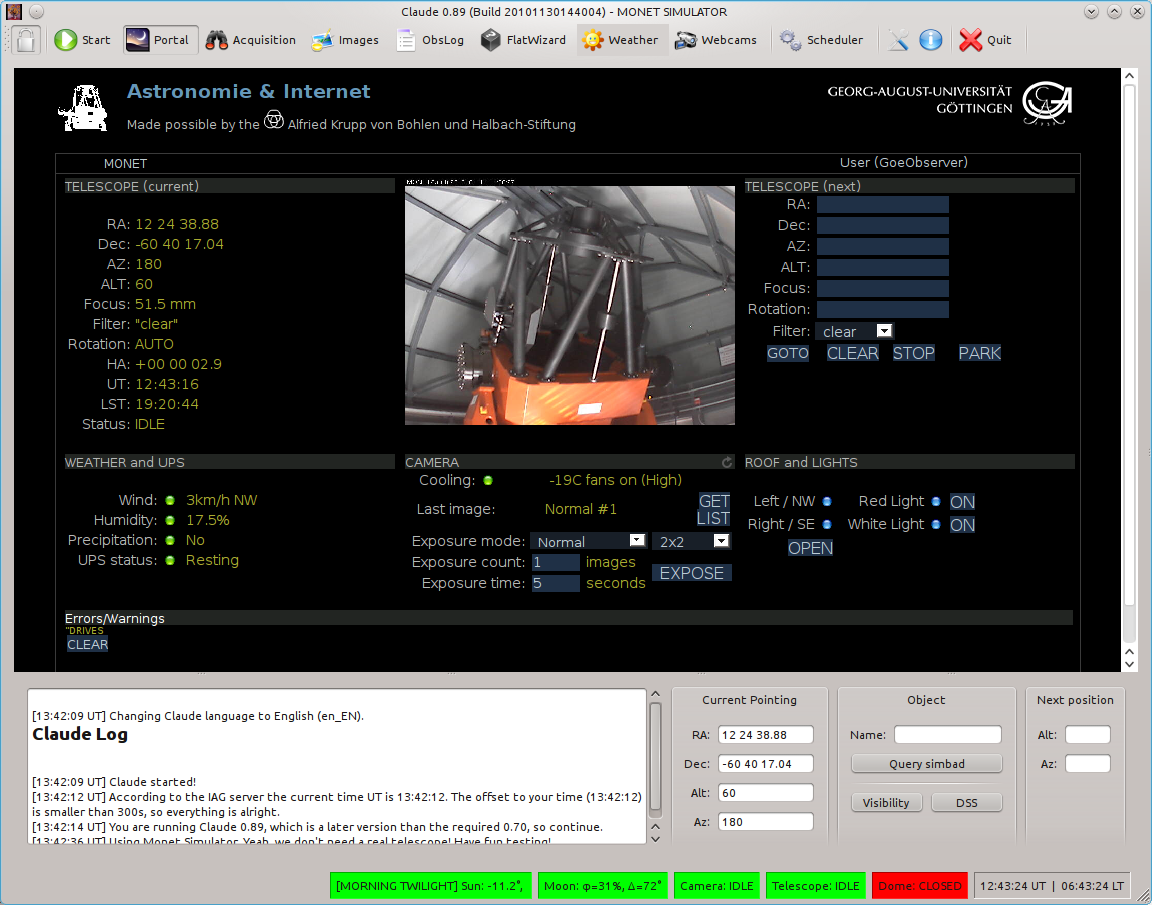}}
  \caption{Two examples of the Monet portal as shown in Claude.}
  \label{fig.Portal}
\end{figure}
To use any of the \emph{Monet} telescopes or the \emph{Simulator}, you have to click on the 
corresponding button. \emph{Claude} now loads the actual \emph{Monet} portal and displays it.
From now on you can use the portal the same way you would if you were not using \emph{Claude}.
All the links and button work (more or less) the same and therefore you can operate the telescope
and take pictures without using any of the advanced features of \emph{Claude}. See fig. \ref{fig.Portal} for
some examples of how the portal is displayed within Claude.

\section{Features added to the portal}
Although the last paragraph stated that the portal as shown inside Claude behaves the same way
as the portal used within a usual web browser, there are some changes that will be described 
her. 

\subsection{Safety precautions}
\begin{figure}
  \centering
  \subfigure[Confirmation for opening the dome.]{\includegraphics[width=0.48\textwidth]{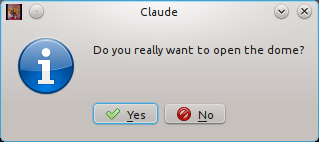}}           
  \subfigure[The sun is still above the horizon.]{\includegraphics[width=0.48\textwidth]{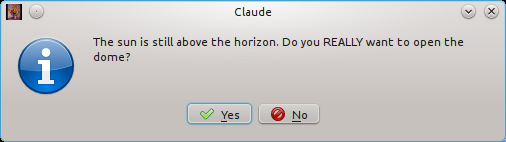}}
  \caption{Claude asks for confirmation when asked to open the dome, especially if the sun is still above the horizon.}
  \label{fig.ConfirmOpenDome}
\end{figure}
Usually when first logging in into the portal and starting an observation, there are two very important buttons:
\textbf{ON} and \textbf{OPEN}. The first one initializes the telescope while the latter one opens the dome.
Unfortunately there is absolutely no safety check implemented in the portal, so you can always click one of this
buttons, even if it is still daylight time and you are facing the risk of exposing the main mirror to the sun. To prevent
this, \emph{Claude} asks you for confirmation, before anything is done. Even during the night a message
box is shown, while during the day a second message box warns you even more explicitly (see fig. \ref{fig.ConfirmOpenDome}).

Furthermore \emph{Claude} will never allow you to move the telescope below 15 degrees altitude.

\subsection{Automatically logging out}
\begin{figure}
  \centering
    \includegraphics{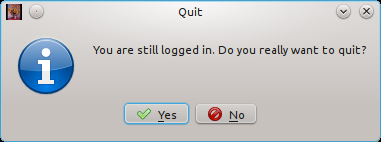}
  \caption{Claude asks user before quitting.}
  \label{fig.ConfirmQuit}
\end{figure}
A very common problem with the \emph{Monet} portal is closing the web browser with the portal by accident. In this
case you cannot login to the portal for some time. \emph{Claude} tries to prevent this by automatically
logging out users on quit. So whenever you have finished your observations and closed down the telescope, just quit
\emph{Claude}, confirm the message box (see fig. \ref{fig.ConfirmQuit}) and you are logged out automatically!

\section{The status section}
\begin{figure}
  \centering
    \includegraphics[width=\textwidth]{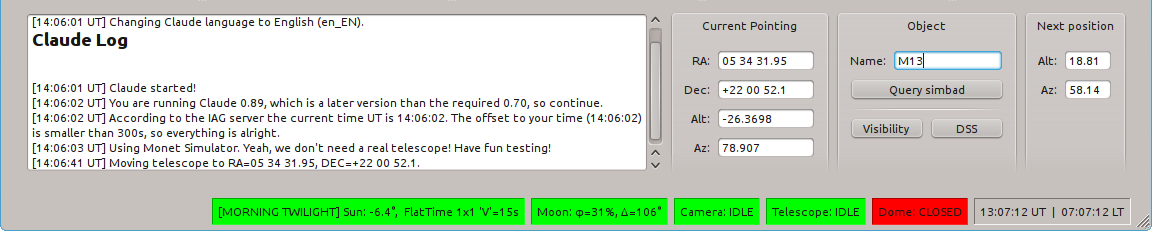}
  \caption{The status section.}
  \label{fig.StatusSection}
\end{figure}
Probably the most obvious feature added to the portal is the status section on the bottom of the
window (see fig.\ref{fig.StatusSection}). 

\subsection{Action log}
On the left there is an action log, where \emph{Claude} prints some information about what it is 
currently doing. The log is automatically stored on disk in the \textbf{Claude directory} that you
specified in the options dialog (see chapter \ref{chap.options}). There is one file for every
night (from noon to noon) with a filename that contains the date of the starting day, e.g.
\texttt{ClaudeLog\_2010\_09\_03.log}.

\subsection{Current pointing}
On its right-hand side the box titled \textbf{Current pointing} shows the current position of the telescope.
Both RA/Dec and Alt/Az are shown and updated regularly. Next to it on the right is the box \textbf{Next
position}. Whenever you enter valid RA/Dec coordinates in the 'TELESCOPE (next)' form 
of the portal, the corresponding Alt/Az position is calculated and displayed here. Therefore you 
can always see how high above the horizon (or if!) a target stands \textbf{before} the telescope is
moved.

\subsection{Object}
\label{subsect.Object}
The remaining box \textbf{Object} was added in order to help you getting your observation started. If you
enter a valid name of an astronomical object, its coordinates are retrieved from \emph{Simbad}
\footnote{\url{http://simbad.u-strasbg.fr/simbad/}} and filled into the \textbf{TELESCOPE (next)} form of the portal.

\begin{figure}
  \centering
    \includegraphics{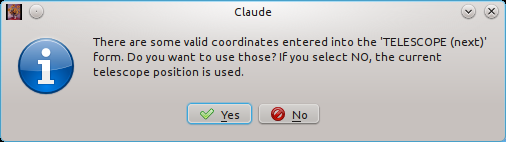}
  \caption{Claude asks user about which coordinates to use.}
  \label{fig.ConfirmWhichCoords}
\end{figure}
\begin{figure}
  \centering
  \subfigure[Visibility chart.]{\label{fig.Visibility}\includegraphics[width=0.48\textwidth]{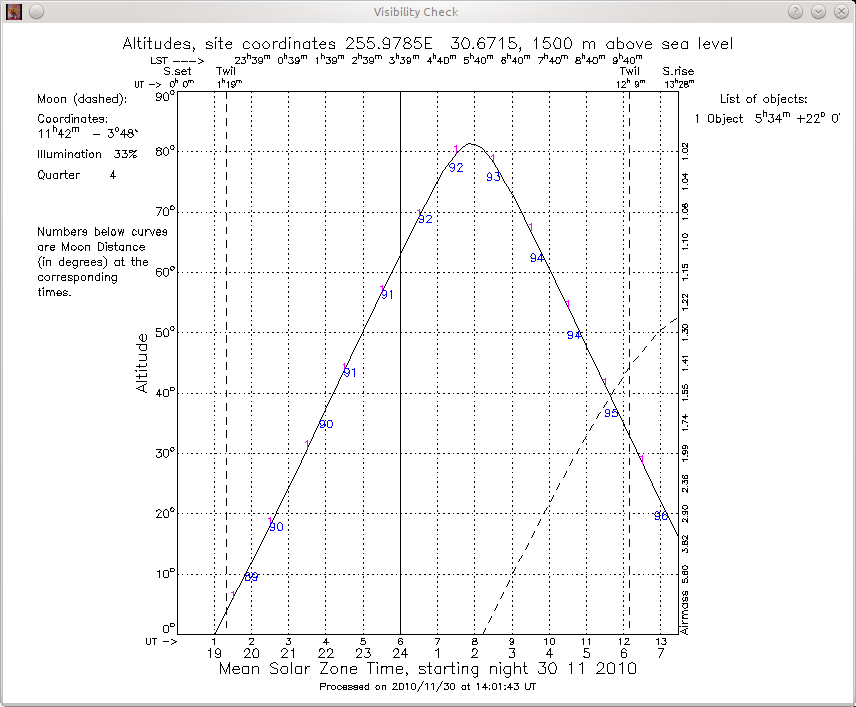}}           
  \subfigure[Finding chart.]{\label{fig.FindingChart}\includegraphics[width=0.48\textwidth]{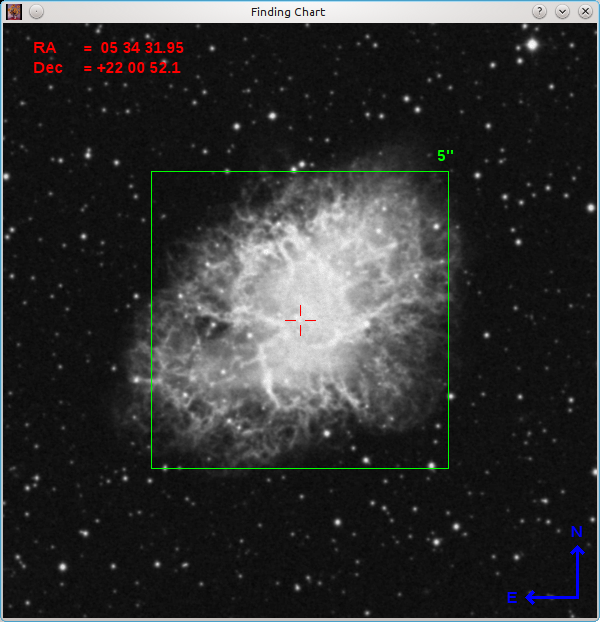}}
  \caption{Visibility and finding charts shown by Claude for a given set of coordinates.}
  \label{fig.VisFindChart}
\end{figure}

For seeing a visibility chart for the current target you can click on \textbf{Visibility}. The graph is
downloaded from the \emph{Isaac Newton Group of Telescopes}\footnote{\url{http://catserver.ing.iac.es/staralt/index.php}}
and displayed in its own window (see fig.\ref{fig.Visibility}). The window can just be closed whenever it is not needed anymore.
In case valid coordinates were entered into the \textbf{TELESCOPE (next)} form of the portal, Claude asks you,
which set of coordinates to use. By clicking \textbf{Yes}, Claude uses the coordinates entered in the portal,
while with \textbf{No} it uses the telescope's current position (see fig. \ref{fig.ConfirmWhichCoords}).

The \textbf{DSS} button works the same way as the \textbf{Visibility} button, but downloads an image for the given coordinates
from the \emph{The STScI Digitized Sky Survey}\footnote{\url{http://archive.stsci.edu/cgi-bin/dss\_form}} and displays
it with some additional information related to \emph{Monet} (see fig. \ref{fig.FindingChart}).

\subsection{Status bar}
On the very bottom of the window there are some additional status information about the telescope. On the far 
right the current time is shown -- both the universal time (UT) and the local time (LT) at the location of
the currently selected telescope.

The three coloured boxes on the left of it show the current status of the camera, the telescope and the dome
(from left to right). A green background colour always indicates everything being okay. For example in fig. 
\ref{fig.StatusSection} you can see that camera and telescope are both idle and that the dome is 
still closed. Here Claude gives immediate feedback whenever an image is exposed or the telescope is moved.
When exposing an image, the number of exposures left is shown together with an estimated time until all
exposures are finished.

\section{Navigating inside Claude}
\begin{figure}
  \centering
    \includegraphics[width=\textwidth]{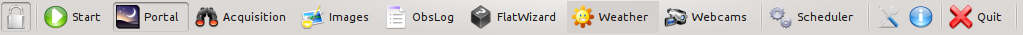}
  \caption{Claude's navigation bar.}
  \label{fig.Navigation}
\end{figure}
\begin{figure}
  \centering
    \includegraphics{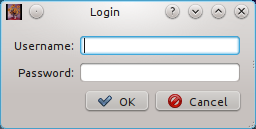}
  \caption{Claude asks user for his username and password.}
  \label{fig.Login}
\end{figure}
On the top of the \emph{Claude} window you can see the navigation bar (see fig. \ref{fig.Navigation}). Nevertheless
the icon on the far left has nothing to do with navigation, but allows you to log in to the portal or log out
by just clicking. The button is enabled only if one of the \emph{Monet} telescopes has been selected before. In case
the you are logging in for the first time, you are asked for your username and password for the portal (see \ref{fig.Login}). 
Both are stored, so you do not have to enter your credentials again until the next restart of Claude.

The next seven buttons in the navigation bar (those between the vertical dashes) allow you to switch between
the different pages provided by \emph{Claude}. The current page is always indicated by a \emph{pushed} button. The two
buttons on the left lead to the Start page as described in section \ref{sec.WrappingPortal} and to the portal.

The other pages will be described in detail later: chapter \ref{chap.siteinformation} deals with \textbf{Weather} and \textbf{Webcams},
while chapter \ref{chap.images} describes how to handle images on the \textbf{Images} and \textbf{Obslog} pages. \textbf{Acquisition} 
(chapter \ref{chap.acquisition}) and \textbf{FlatWizard} (chapter \ref{chap.flatwizard}) will also be explained in extra
chapters of this manual.

Most of the pages are being enabled only after chosing a telescope site or even only after activating the telescope.

The next three buttons do not belong to another page, but open up new windows with (from left to right) the \textbf{Scheduler}
(chapter \ref{chap.scheduler}), the options and some informations about \emph{Claude} itself (both chapter \ref{chap.options}).

The last button quits Claude.

%% file: siteinformation.tex
\chapter{Telescope site information}
\label{chap.siteinformation}

\section{Weather}
\begin{figure}
  \centering
    \includegraphics[width=\textwidth]{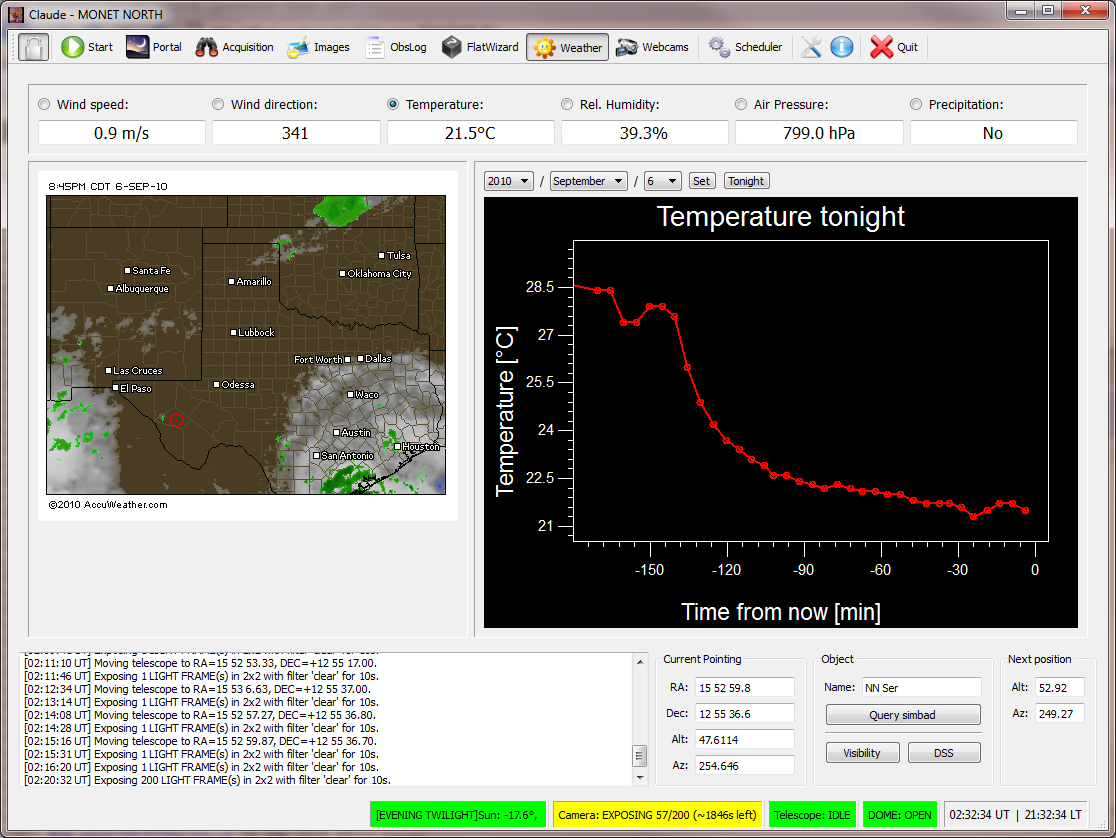}
  \caption{Weather information from the currently selected telescope site.}
  \label{fig.Weather}
\end{figure}
\emph{Claude} automatically gathers weather information from the telescope site and displays it in a convenient way 
(see fig. \ref{fig.Weather}). All the weather data is automatically stored on disk in the \textbf{Claude directory} that you
specified in the options dialog (see chapter \ref{chap.options}). There is one file for every
night (from noon to noon) with a filename that contains the site location and the date of the starting day, e.g.
\texttt{Weather\_MonetNorth\_2010\_11\_30-01.log}.

\subsection{Weather map}
A video containing a map with the cloud coverage and detected rain over the last couple of hours is
updated in a regular interval and shown on the left of the page. The data comes from:
\begin{itemize}
	\item \emph{Monet/North}: Accuweather\footnote{\url{http://sirocco.accuweather.com/sat\_mosaic\_400x300\_public/rs/isarTXW.gif}}.
	\item \emph{Monet/South}: not supported yet.
	\item \emph{Monet/Sim}: same as \emph{Monet/North}.
\end{itemize}
On the Accuweather maps, clouds are indicated in white and rain in different shades of green, i.e. the greener the more rain.

\subsection{Current weather}
On the top of the page there is a list of current weather data. The data is being updated regularly from:
\begin{itemize}
	\item \emph{Monet/North}: McDonald Observatory Current Weather\footnote{\url{http://198.214.229.50/cgi-bin/obs\_sup/latest\_5min.cgi}}.
	\item \emph{Monet/South}: not supported yet.
	\item \emph{Monet/Sim}: same as \emph{Monet/North}.
\end{itemize}
Next to the labels for the text boxes that show the actual numbers, there are radio boxes that control the plot
that is described next.

\subsection{Plot}
Quite often the trend for e.g. the temperature or the humidity is of more interest than the current value itself.
Therefore \emph{Claude} plots the selected weather data for the last up to 180 minutes. The selection is done using
the radio boxes at the top of the page as described in the last paragraph.

Instead of plotting the weather data of the last three hours, one can select another night from the drop-down
lists above the plot. After selecting a year, month and day (of the start of the night) and clicking 
\textbf{Set}, \emph{Claude} loads the corresponding file from disk and plots the data for the entire night. For this of course 
there must be a log file for that specific night, i.e. you must have been observing with Claude back then.

\section{Webcams}
\begin{figure}
  \centering
    \includegraphics[width=\textwidth]{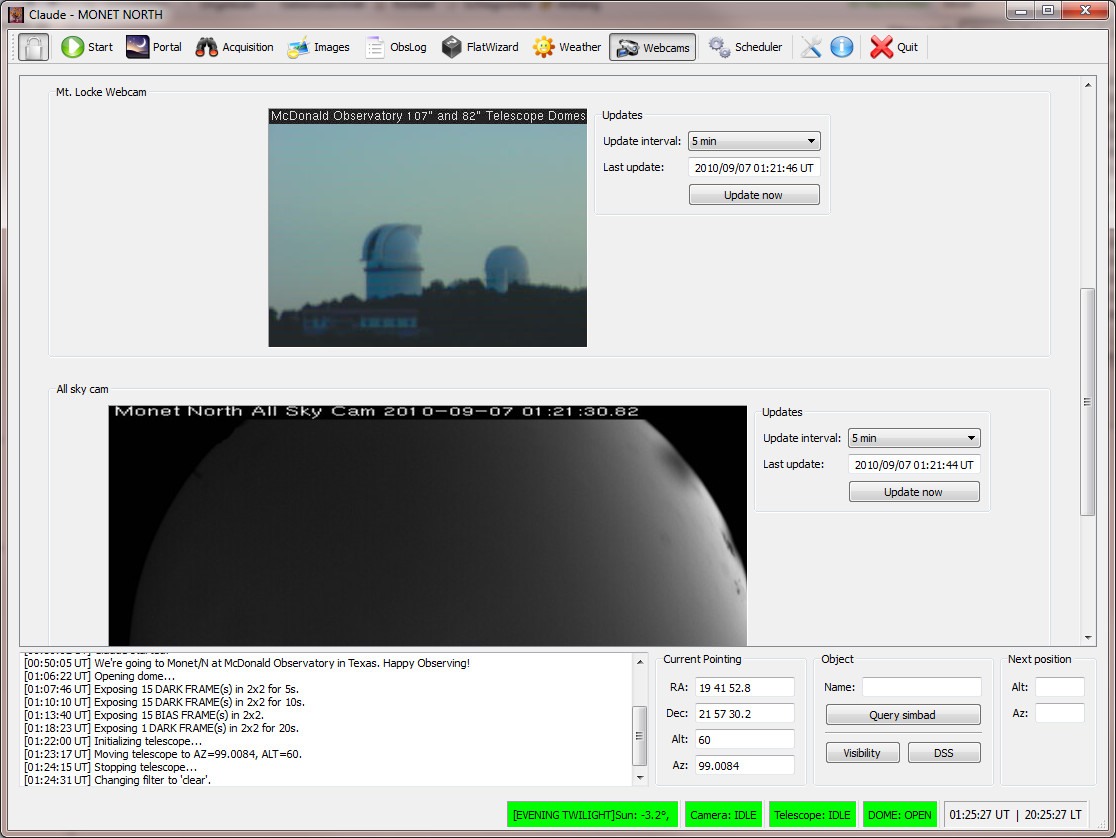}
  \caption{Webcams showing the selected telescope site.}
  \label{fig.Webcams}
\end{figure}
On the \textbf{Webcams} page a predefined list of webcam images are shown. For \emph{Monet/North} these are:
\begin{itemize}
	\item Inside webcam\footnote{\url{http://monet-cam3.as.utexas.edu/axis-cgi/jpg/image.cgi?resolution=320x240}}
  \item Outside webcam\footnote{\url{http://monet-cam.as.utexas.edu/axis-cgi/jpg/image.cgi?resolution=320x240}}
  \item Mt. Locke Webcam\footnote{\url{http://mcdcam-dimm.as.utexas.edu/axis-cgi/jpg/image.cgi?resolution=320x240}}
  \item All sky cam\footnote{\url{http://http://198.214.229.115/jpg/1/image.jpg}}
\end{itemize}
For \emph{Monet/Sim} the webcams are the same; \emph{Monet/South} is not supported yet.

Next to the webcam images there is a listbox for changin the update behaviour of the images. The supported update intervals range from
'Never' to '30 min', the default is '15 min'. Usually during the night you can disable all webcams
except the All sky cam, since they will be all black anyway.

The text box below shows the time of the last update and a click on \textbf{Update now} overrides the selected
update interval and updates the webcam image immediately.

%% file: images.tex
\chapter{Images}
\label{chap.images}

\section{Online image summaries}
\begin{figure}
  \centering
    \includegraphics[width=\textwidth]{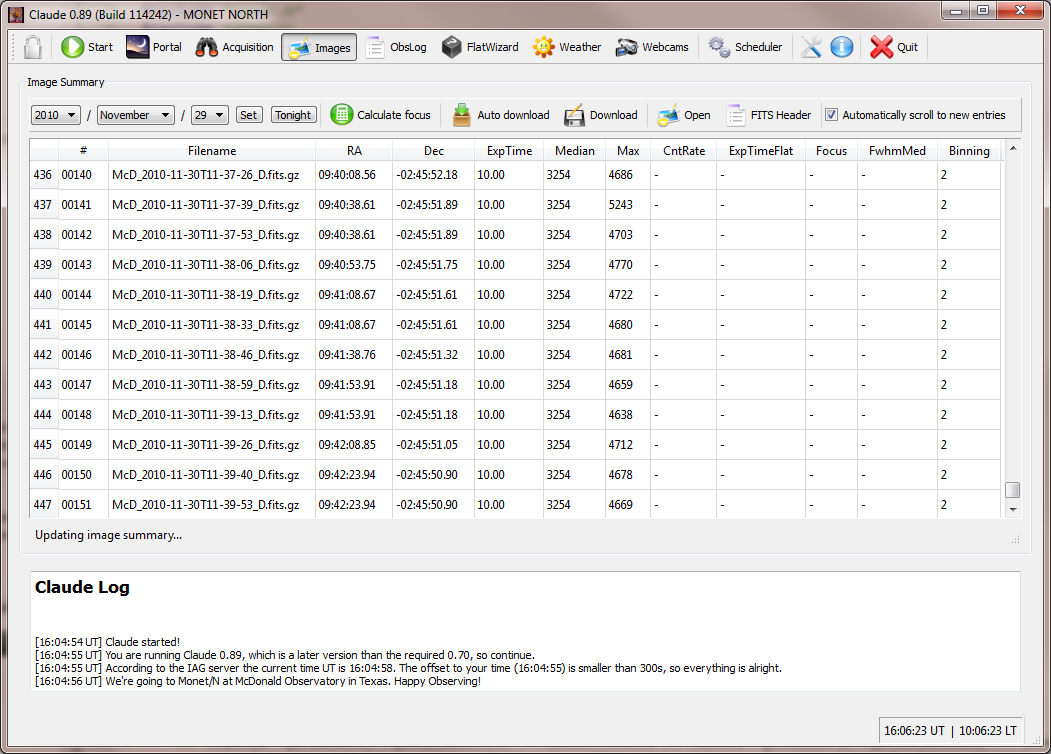}
  \caption{Image summary of selected night.}
  \label{fig.ImageSummary}
\end{figure}
A list of images taken during a night is stored on servers at the telescope sites and are publicly available. See 
\url{http://alfred.as.utexas.edu/~mhunder/nightlog_2010_12_09-10.log} for an example for \emph{Monet/North}. There
also is a log available with only the latest fifteen images (\url{http://alfred.as.utexas.edu/~mhunder/median_counts.txt}), 
which is easier to handle during observations. Both log files share the same format. The columns are (from left to right):
\begin{itemize}
 \item A counting number (1) and the filename (2), which may be truncated due to the limited space in the log,
 \item Right Ascension (3) and Declination (4) of the telescope during the exposure,
 \item Exposure time (5),
 \item median (6) and maximum (7) counts in image,
 \item exposure time estimation for sky flats (9),
 \item current focus value of telescope (10) and median (11), mean (12) and standard deviation (13) of the FWHMs of the stars 
       in the image, as determined using SExtractor\footnote{\url{http://www.astromatic.net/software/sextractor}},
 \item binning of the image (14) and airmass during the exposure (15).
\end{itemize}

\section{List of images}
Using \emph{Claude}, manually keeping track of these logs is unnecessary. Instead, \emph{Claude} can display the night logs
of all previous nights on its \textbf{Images} page (see fig. \ref{fig.ImageSummary}). The list is updated automatically 
whenever a new image has been taken. However, this can take up to thirty seconds after an exposure is finished.
Previous nights can be selected by chosing the date at the beginning of the night in the drop-down lists on the top of the page
and clicking \textbf{Set}. \textbf{Tonight} always takes you back to the current night.

The automatic behaviour of \emph{Claude} is to automatically scroll down to the latest entry in the list when a new
image has been found. Since this can become quite annoying when looking through older images, it can be turned off using the
check-box on the far right.

In the list multiple images can be selected at once following operating system standards. In Windows and
Linux these would be for example:
\begin{itemize}
 \item Select one image, hold down SHIFT and select another one. All the images in between (including the selected ones)
       will be selected.
 \item Hold down the CTRL key to select multiple images.
 \item Click on an image without holding any key, so that only that image will be selected.
\end{itemize}

\section{Telescope focus}
Although \emph{Claude} can focus the telescope automatically (see section \ref{subsect.focus_telescope}), one sometimes wants 
to calculate the focus manually from a set of images. This can be done by selecting all the images of a focus series and clicking
on \textbf{Calculate focus}. \emph{Claude} will show a plot of the FWHMs of the stars in the images over the focus value together
with an estimated best focus values as determined from a hyperbolic fit (see fig. \ref{fig.Focus}).

\section{Downloading images}
\emph{Claude} offers two methods for downloading images from the telescope sites: a manual and an automatic one.
\begin{itemize}
 \item \textbf{Manual mode}: Select a couple of images from the list and click \textbf{Download}. \emph{Claude} will
       ask you for a directory to save the images to and start the download.
 \item \textbf{Automatic mode}: The \textbf{Auto download} button behaves like a check-box, i.e. it triggers when
       clicked. To use this feature, a download directory must have been set (see chapter \ref{chap.options}).
       Every new image will now be downloaded to this directory.
\end{itemize}

In the options dialog (see chapter \ref{chap.options}) there also is an option for decompressing the gzipped images
automatically into the download directory. If you are using a FITS viewer that cannot handle compressed images,
this makes it much more comfortable to handle \emph{Monet} images.

Also in the options you can set the maximum number of files to download simultaneously. To see which images are
still in the queue to be downloaded, hit F8.

\section{Opening an image}
If a FITS viewer is defined (see chapter \ref{chap.options}), clicking on \textbf{Open} after selecting an image from
the list opens this viewer with the selected image.

\section{FITS headers}
\begin{figure}
  \centering
  \subfigure[Plot of FWHMs over focus value for focussing telescope.]{\label{fig.Focus}\includegraphics[width=0.48\textwidth]{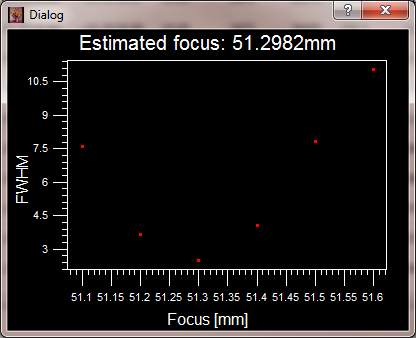}}           
  \subfigure[List of all FITS headers for a given file.]{\label{fig.FitsHeader}\includegraphics[width=0.48\textwidth]{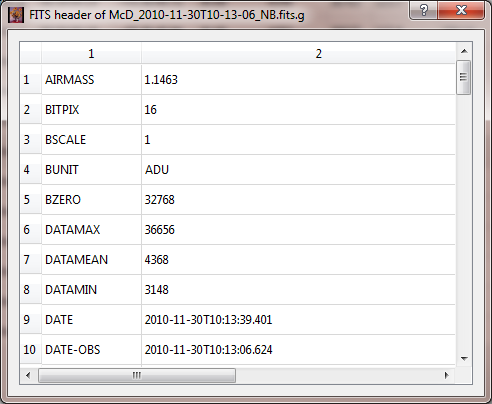}}
  \caption{Two of the features that can be activated from the image summary.}
\end{figure}
If you are interested in the FITS headers of an image, you can select it in the list and click on \textbf{FITS Header}.
A new window will open up with all the key/value pairs in the file (see fig. \ref{fig.FitsHeader}).

%% file: acquisition.tex
\chapter{Acquisition}
\label{chap.acquisition}
\begin{figure}
  \centering
    \includegraphics[width=\textwidth]{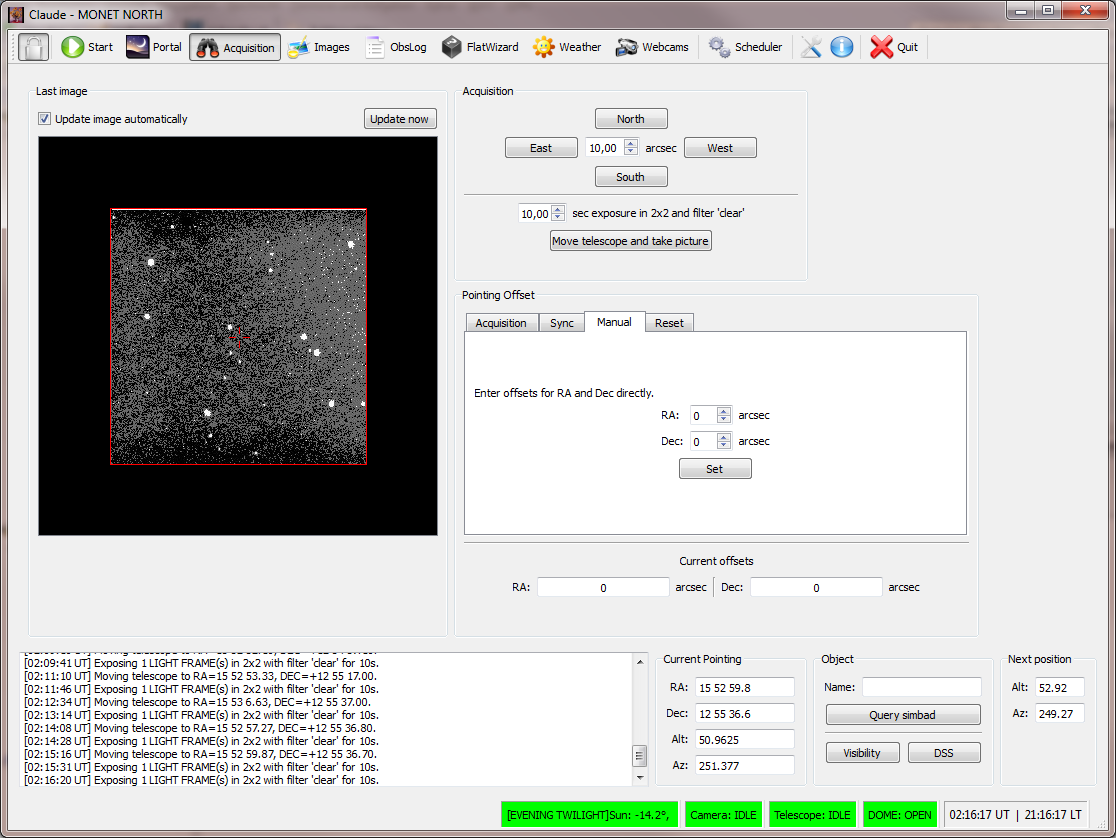}
  \caption{Claude's acquisition page.}
  \label{fig.Acquisition}
\end{figure}

\section{Last image taken}
When operating a telescope on site, it is usually easy to have a look at the last image taken. For remote telescopes like 
\emph{Monet} this is quite different, since the portal page does not show it, so you have to download it from the server
and open it in a FITS viewer manually.

But still this solution is far from perfect, since the images have to be downloaded over not-so-fast internet
connections, so it can take a little longer than usual to be able to see an image just taken. To make life easier
(and observations quicker), a preview image of the last image is provided on the server. For \emph{Monet/North} e.g. this
can be found at \url{http://alfred.as.utexas.edu/~mhunder/preview.jpg}. It is a downsized and compressed JPEG image,
which is small enough in size to be downloaded in a very short amount of time.

\emph{Claude} can also download and display this image (see fig. \ref{fig.Acquisition}). The check-box above the
image controls, whether a new image is downloaded automatically or only on a click on the \textbf{Update now}
button.

The image has the same orientation as the finding chart image downloaded from DSS (see section \ref{subsect.Object}), so you
can always compare both in order to get a feeling about where the telescope currently points to and where to move next.

\section{Finding a target}
Especially for extended targets, the coordinates provided by e.g. \emph{Simbad} are often only a first guess for the final
coordinates and need to be refined by taking pictures and moving the telescope subsequently over and over again. Using the 
preview image, you can of course use the \emph{Monet} portal to move the telescope, but \emph{Claude} offers a far more efficient 
way to do this.

Using the four buttons in the \textbf{Acquisition} box, you can move the red rectangle overlaying the preview image. The step
size can be varied by the text box in the middle -- the default is 10 arcseconds.

Once the red rectangle covers the area you are interested in, there are two possibilities offered by the two buttons below:
\begin{itemize}
 \item \textbf{Move telescope}: Just move the telescope. This is useful e.g. when doing time-series of an object that
       tends to move out of the field of view. In that case you can move the red rectangle to the desired position and
       click on \textbf{Move telescope}. As a result one or two images might be blurred, but the time-series can be
       continued without an interruption.
 \item \textbf{Move telescope \& take picture}: While still in the process of finding the best pointing for the telescope,
       you can use this button as a real \textbf{Acquisition mode}. The telescope is moved according to the red rectangle
       and an image in clear is taken with an exposure time given in the text box above.
\end{itemize}

\section{Pointing offsets}
\begin{figure}
  \centering
  \subfigure[Use values from acquisition.]{\includegraphics[width=0.48\textwidth]{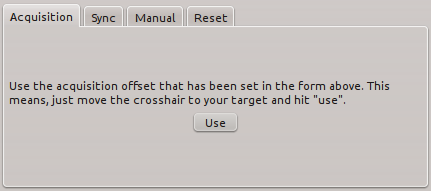}}           
  \subfigure[Synchronize with given set of coordinates.]{\includegraphics[width=0.48\textwidth]{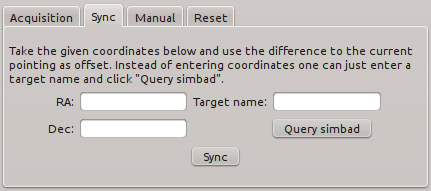}} \\
  \subfigure[Enter offsets manually.]{\includegraphics[width=0.48\textwidth]{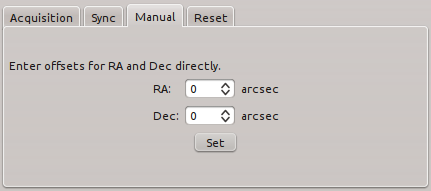}}           
  \subfigure[Reset offsets.]{\includegraphics[width=0.48\textwidth]{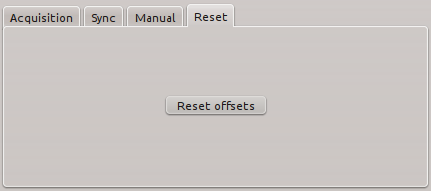}} \\
  \caption{Setting a pointing offset in \emph{Claude}.}
  \label{fig.PointingOffset}
\end{figure}
Sometimes it can happen that the poining of the \emph{Monet} telescopes is not quite perfect. When operating the telescope manually
this can be corrected by moving the telescope to its desired position before starting the observations, e.g. by using
\emph{Claude's} Acquisiton mode (see last section). This becomes cumbersome when observing multiple targets and quite
impossible when operating in automatic mode (see chapter \ref{chap.scheduler}).

If the targets to observe are all in a confined area on the sky, a constant pointing offset is a good approximation
for the real correction for the pointing model. This works e.g. for observations within the galactic bulge.

\emph{Claude} offers several ways to set a constant offset like this:
\begin{itemize}
 \item \textbf{From acquisition}: You can use the red rectangle from the acquisiton mode for a pointing offset. Therefore,
       after moving the rectangle, instead of moving the telescope, you can click the \textbf{Use} button here and the offset
       of the rectangle to its original position is used as pointing offset.
 \item \textbf{Synchronize with coordinates}: After really moving the object, which already should be at the center of
       the preview image, to the center (i.e. under the red cross), you can enter its coordinates again here (or query
       \emph{Simbad} for this purpose) and click on \textbf{Sync}. The difference between the actual and the given coordinates will
       be used as pointing offset.
 \item \textbf{Manual}: Of course you can enter pointing offset for RA and Dec manually.
 \item \textbf{Reset}: Resets both offsets to zero.
\end{itemize}

The current pointing offsets are shown at the bottom of the \textbf{Pointing offset} box (see fig. \ref{fig.Acquisition}) and will
be added to every pair of coordinates the telescope is moving to.

%% file: flatwizard.tex
\chapter{Flat-field wizard}
\label{chap.flatwizard}

The flat-field wizard is still work in progress. The documentation for it will be added as soon as it is finished.

\begin{figure}
  \centering
    \includegraphics[width=\textwidth]{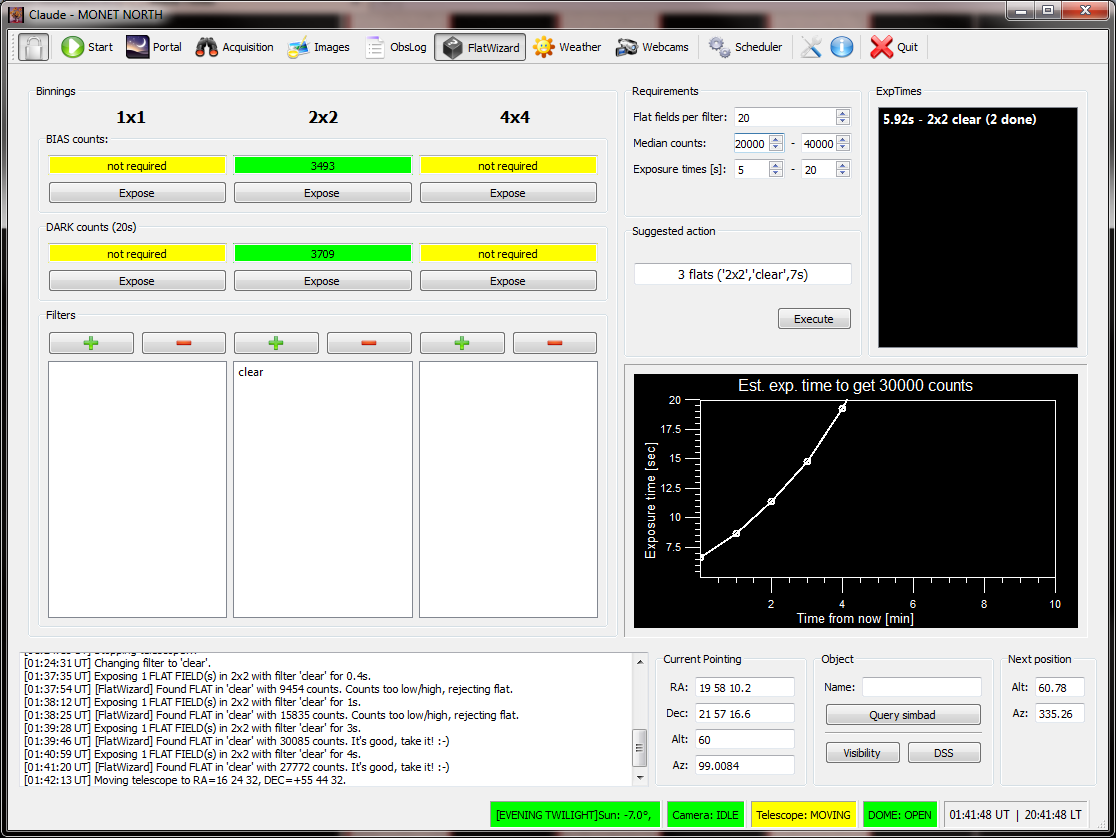}
  \caption{Claude's flat-field wizard.}
  \label{fig.FlatFielding}
\end{figure}

%% file: scheduler.tex
\chapter{Scheduler}
\label{chap.scheduler}

\section{Managing tasks}
\begin{figure}
  \centering
    \includegraphics{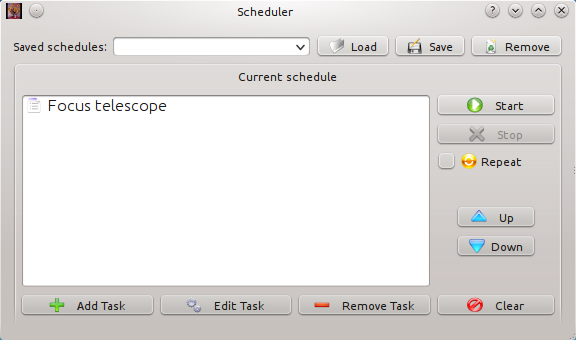}
  \caption{The task scheduler within \emph{Claude}.}
  \label{fig.Scheduler}
\end{figure}
\emph{Claude} offers a way for automating different tasks that do not need user interaction. This is not comparable
to the planned robotic mode for the \emph{Monet} telescopes, but can make an observer's life easier in several
cases. Single tasks (e.g. moving the telescope) can be run, but also a list of tasks can be defined (e.g. move
telescope and take picture).

A new task can be added by clicking on \textbf{Add task} and selecting it from the popup menu. The options dialog
for the selected task will show up and upon accept the task will be added to the list.

Once in a list, a task's options dialog can be opened again by selecting it and clicking \textbf{Edit task}. A task
can also be removed using the \textbf{Remove task} button. The whole schedule can be emptied with the \textbf{Clear}
button. For changing the order of tasks, you can use the arrow buttons on the right, which move the selected task
up or down in the list.

\section{Running the scheduler}
The \textbf{Start} button starts the scheduler with the selected task. If no task is selected, it starts with the
first one in the list. All the tasks are run subsequently till the end of the list, unless the \textbf{Repeat}
checkbox is checked. In that case, after finishing the last task, the scheduler starts again with the first task
in the list. The scheduler can be stopped by clicking \textbf{Stop}. 

\textbf{ATTENTION}: This button \textbf{only} stops
the scheduler and \textbf{not} any running telescope or camera operation (e.g. moving telescope, taking picture)!
Those have to be stopped manually in the portal!

While the scheduler is running, the list of tasks cannot be changed, i.e. no tasks can be added/removed/edited.

\section{Loading and saving schedules}
Schedules can be saved on disk and loaded again later. In order to save it, you have to provide a name in the textbox
at the top and click \textbf{Save}. Afterwards the saved schedule will show up in the dropdown list associated
with that textbox. Selecting it and clicking \textbf{Load} will load it again. Saved schedules can be removed by
selecting them and using the \textbf{Remove} button.

\section{Tasks}
There are some pre-defined tasks that will be explained in detail in the following.

\subsection{Move telescope}
\begin{figure}
  \centering
    \includegraphics{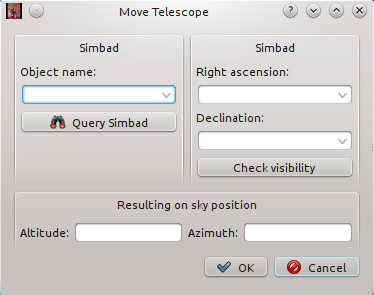}
  \caption{The task for moving the telescope.}
  \label{fig.TaskMoveTelescope}
\end{figure}
One of the most simple tasks just moves the telescope to a new RA/Dec position. Those coordinates can either be
entered manually using the two textboxes on the right (see fig. \ref{fig.TaskMoveTelescope}), or an object's
name can be entered in the \textbf{Simbad} box on the left. A click on \textbf{Query simbad} will then let \emph{Claude} try
to fetch the object's coordinates from \emph{Simbad}. Both textboxes for RA and Dec also act as dropdown lists and will
show all the coordinates that have previously been used.

At the bottom of the dialog, the current Alt/Az position of the given RA/Dec coordinates are shown. Please be aware,
that \emph{Claude} checks the coordinates again before the task is executed and will never move the telescope
below an altitude of 15 degrees.

The \textbf{Check visibility} button will produce a visibility chart as described in section \ref{subsect.Object}.

\subsection{Expose image}
\begin{figure}
  \centering
    \includegraphics{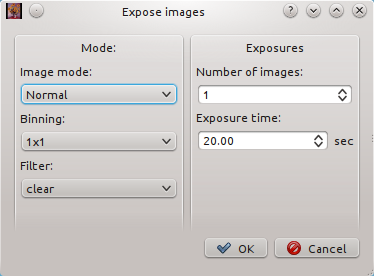}
  \caption{The task for exposing an image.}
  \label{fig.TaskExposeImage}
\end{figure}
Another basic task allows to automate exposing images (see fig. \ref{fig.TaskExposeImage}). This is e.g. useful
for taking RGB images. Instead of manually changing the filter and exposing images for every color, one can use a
set of this tasks to accomplish it.

On the left, the image mode ('Normal', 'Bias', 'Dark') can be changed together with the binning (currently 1x1, 
2x2 and 4x4) and the filter (only for Normal frames!). Be aware that a filter change always takes some time, 
so a filter cycle of RRRGGGBBB is more efficient than RGBRGBRGB.

The number of images to expose and the exposure time (only for Normal and Dark frames!) can be edited on the right.

\subsection{Focus telescope}
\label{subsect.focus_telescope}
\begin{figure}
  \centering
    \includegraphics{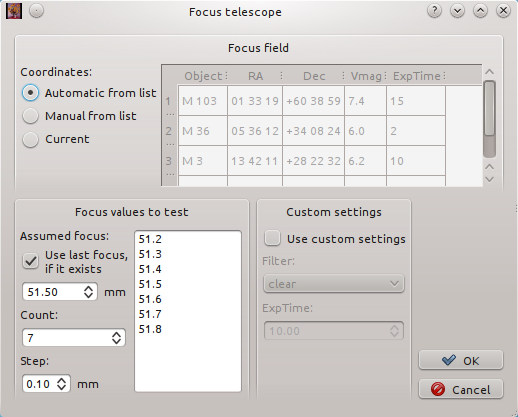}
  \caption{Focussing the telescope using \emph{Claude}.}
  \label{fig.TaskFocusTelescope}
\end{figure}
Focussing can be one of the most cumbersome and annyoing tasks when operating a telescope. Serendipitously \emph{Claude}
can do almost all the work necessary. See fig. \ref{fig.TaskFocusTelescope} for the options dialog for \emph{Claude's} task
to focus the telescope.

At first you have to select, which field on the sky to use. There is a list of pre-defined objects in the list on the upper
right. The default setting is to let \emph{Claude} decide automatically, which of these fields to use. It is always that object 
selected that is currently highest up in the sky. Alternatively you can select an object from the list manually. Of course
it is also possible to focus on a completely different target (i.e. the science target). In that case you should move the
telescope to that coordinate manually (or use the \textbf{Move telescope} task) and select \textbf{Current} from the
options on the left.

In the box on the lower left the list of focus values that should be tested are defined. \emph{Claude} automatically fills
the \textbf{Assumed focus} with the focus value currently set, since that is presumably at least a good first guess. Of course
the value can be edited manually. The checkbox above allows \emph{Claude} to change this value again, when the task
has been started. If a focus value has been determined before, that one is used. This is especially useful, if one wants
to refocus the telescope multiple times during one night. Below a step size and the number of focus values to test can be 
changed. Experience showed that seven steps with a step size of 0.1mm yield good results. The listbox on the right shows a list 
of focus values that will be tested.

Usually \emph{Claude} always focusses the telescope in the 'Clear' filter. Since the focus is just a function of filter
width, the values for the other filters can be calculated from the known widths. Alternatively one can force \emph{Claude}
to focus using a different filter and different exposure times. To do this, just check \textbf{Use custom settings} and
edit the values manually.

When the task is run, the telescope is moved to a new target (if defined), steps through all the defined focus values and
takes an image. Afterwars a hyperbola fit is done in order to calculate the best focus value. Beware that this task will
\textbf{never} actually set the focus value. See the next task for this.

\subsection{Set telescope focus}
\begin{figure}
  \centering
    \includegraphics{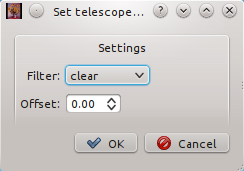}
  \caption{Setting the focus for a given filter.}
  \label{fig.TaskSetTelescopeFocus}
\end{figure}
After determining the telescope focus (see last task), you usually wants to actually set it. In order to do this, the
corresponding task asks for the used filter and an optional offset (see fig. \ref{fig.TaskSetTelescopeFocus}).

As described in the last section, the telescope is always focussed using the 'Clear' filter. This tasks needs to know
the actual filter used in order to calculate the best focus value. In some cases it might be necessary to take unfocussed
images (i.e. with targets too bright). For this an offset can be defined that will always be added the calculated 
focus value. A value of $0.0$ (i.e. no offset) always yields the best focus value to be set.

\section{Writing your own task}
It is possible to write your own tasks using a Javascript-like language. If you are interested in this, please contact
the author for further instructions.

%% file: options.tex
\chapter{Options}
\label{chap.options}

\section{Language}
\begin{figure}
  \centering
    \includegraphics{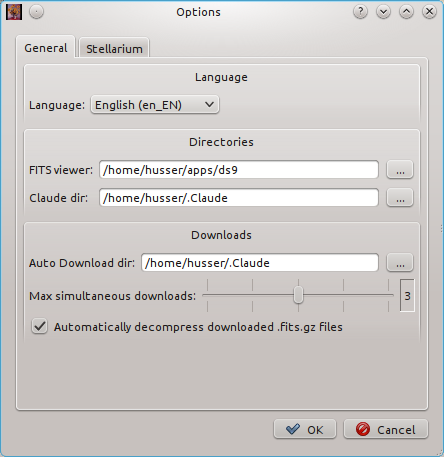}
  \caption{General settings in \emph{Claude's} options dialog.}
  \label{fig.OptionsGeneral}
\end{figure}
\emph{Claude's} default language is English, nevertheless that can be changed using \emph{Claude's} 
options dialog (see fig. \ref{fig.OptionsGeneral}), which can be accessed by clicking the tool shaped 
icon in the tool bar.

Currently \emph{Claude} provides only one more language: German. If anyone is interested in creating
translations for other languages, please contact the author.

\section{Directories}
\emph{Claude} needs one directory, where it can write to in order to store logs, schedules and other stuff.
This directory is usually refereed to as \textbf{Claude dir}. It is absolutely necessary and
\emph{Claude} will not run properly without it being set.

Furthermore you can set the path to a FITS viewer (like DS9, ImageJ, ...) here. This is then used to open
files from the image summary (see chapter \ref{chap.images}).

\section{Downloads}
For using the \emph{Auto download} feature of the image summary, you also need to set the \textbf{Auto
download dir} in the options dialog. This directory is where all the images are downloaded to
automatically.

Since both \emph{Monet/North} and \emph{Monet/South} do not have stunningly fast internet connections,
it is always a good idea to limit the maximum number of simultaneous image downloads. This can be done
using the slider provided in the options dialog. If more than the given number of downloads are
requested, they are put in a queue, which can be displayed by pressing the F8 key.

In order to limit the amount of data that needs to be transferred, the images are stored on the server
in a compressed archive. The checkbox at the bottom lets \emph{Claude} decompress those archives
automatically after the download.

\section{Stellarium}
\begin{figure}
  \centering
    \includegraphics{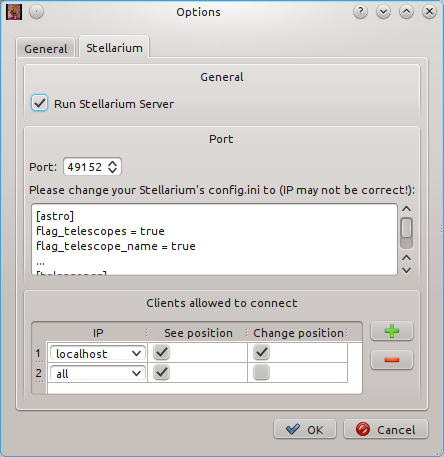}
  \caption{Stellarium settings in \emph{Claude's} options dialog.}
  \label{fig.OptionsStellarium}
\end{figure}
A feature especially interesting for schools and groups of students is the possibility of controlling \emph{Claude}
(and therefore the \emph{Monet} telescopes) remotely using \emph{Stellarium}\footnote{\url{http://www.stellarium.org/}},
which is a free open source planetarium available for Windows, Linux and MacOS.

When using this feature, \emph{Stellarium} always shows the current position of the connected \emph{Monet} telescope and even
allows its user to move the telescope by clicking on an object. Please see the \emph{Stellarium} manual for further details.

In order to use \emph{Claude's} \emph{Stellarium} server, it must be activated in the options dialog (see fig. \ref{fig.OptionsStellarium}).
Futhermore a TCP port must be specified, which is used to communicate with \emph{Stellarium}. If you do not know, what this is,
just do not touch it! For older versions of \emph{Stellarium} a configuration file needed to be changed and the lines in the
text box below added. For more recent versions this can be done using \emph{Stellarium}'s \textbf{Remote plugin}. Please see the
manual for details.

Since we do not want to allow every \emph{Stellarium} client in the local network to connect to \emph{Claude}, a list of allowed
clients can be edited below. The default setting is that every \emph{Stellarium} client can see the current position of the
\emph{Monet} telescope (whatever, no harm done), but only a \emph{Stellarium} running on the same machine as \emph{Claude}
can request the telescope to move. This options probably need to be adapted in order fit the needs of a work group using
this feature. It might e.g. be a good idea to run \emph{Claude} on one computer and use another one to control it
via \emph{Stellarium}.

Whenever \emph{Claude} receives a request from one of the \emph{Stellarium} clients to move the telescope, it asks for confirmation 
before it actually does anything. This way nobody can move the telescope without approval of the observer that runs
\emph{Claude}.